# Thermodynamics Beyond Equilibrium in Coherent Quantum Systems

Md Manirul Ali[1*] and Po-Wen Chen[2]

[1]Centre for Quantum Science and Technology, Chennai Institute of Technology, Sarathy Nagar, Kundrathur, Chennai, 600069, India.
[2]Department of Physics, National Atomic Research Institute, Taoyuan, 325207, Taiwan.

*Corresponding author(s). E-mail(s): manirul@citchennai.net;
Contributing authors: powen@nari.org.tw;

**Abstract**

A central open question in quantum thermodynamics is the identification of thermodynamic entropy away from equilibrium. While thermodynamic entropy coincides with the von Neumann entropy of a Gibbs state at thermal equilibrium, coherent nonequilibrium evolution generates quantum coherence whose thermodynamic role remains unclear. Here we show that the von Neumann entropy does not, in general, represent the thermodynamic entropy governing heat exchange in coherent nonequilibrium quantum systems. By integrating the resource theory of quantum coherence with thermodynamic principles, we develop a general framework for nonequilibrium quantum thermodynamics. We identify the thermodynamic entropy with the energy entropy associated with energy-measurement statistics and introduce a dynamical temperature that extends the equilibrium concept of temperature to nonequilibrium states. Within this framework, we obtain an exact entropy-balance relation that separates the entropy flux associated with heat exchange from the coherence-loss contribution to entropy production, thereby establishing an operational connection between quantum coherence, heat flow, and total entropy change. Unlike conventional approaches based on equilibrium reference states or stationary Gibbs ensembles, the present theory is formulated entirely in terms of instantaneous dynamical quantities and remains valid throughout transient nonequilibrium evolution. Temperature, work, heat, and free energy therefore acquire consistent dynamical definitions far from equilibrium. Applying the formalism to an exactly solvable open quantum system, we demonstrate the dynamical emergence of equilibrium thermodynamics in the weak-coupling limit. Our results provide a unified framework for coherent nonequilibrium quantum thermodynamics and suggest new opportunities for controlling heat dissipation through coherence engineering in quantum thermal technologies.

# 1 Introduction

Thermodynamics provides one of the most successful and universal descriptions of physical systems, establishing fundamental relations between energy, heat, work, entropy, and temperature. While classical thermodynamics was originally developed for macroscopic systems in thermal equilibrium [1, 2], recent advances in quantum technologies have stimulated enormous interest in extending thermodynamic concepts to microscopic quantum systems operating far from equilibrium [3–9]. Quantum thermal machines [10–15], quantum batteries [16–20], and thermoelectric transport [21–25] devices typically operate under nonequilibrium conditions where quantum coherence plays an increasingly important role. Understanding the role of quantum coherence in thermodynamics has therefore emerged as one of the central challenges of modern quantum science.

Quantum coherence represents one of the most distinctive resources of quantum systems and has recently been formalized within a rigorous resource-theoretic framework [26, 27]. Quantum coherence has been shown to influence work extraction in quantum heat engines, energy storage performance in quantum batteries, and the implementation of thermal operations in quantum regime [28–34]. Nevertheless, its precise role during irreversible nonequilibrium evolution remains incompletely understood. In particular, it is important to clarify how coherence contributes to entropy generation and heat dissipation in open quantum systems.

A fundamental open question in nonequilibrium quantum thermodynamics concerns the identification of thermodynamic entropy away from equilibrium. In thermal equilibrium, the Gibbs state is diagonal in the energy eigenbasis and the thermodynamic entropy coincides with the von Neumann entropy [35] of the quantum state. Consequently, the von Neumann entropy is frequently employed in analyses of nonequilibrium quantum processes and quantum thermal devices. However, coherent nonequilibrium dynamics inevitably generate off-diagonal density matrix elements in the energy eigenbasis. This raises an important conceptual question: does the von Neumann entropy continue to represent the thermodynamic entropy governing heat exchange and thermodynamic irreversibility in presence of quantum coherence? Resolving this issue remains one of the central open problems in nonequilibrium quantum thermodynamics [8, 36–39].

Finding the appropriate definition of thermodynamic entropy away from equilibrium is also closely connected to the problem of the microscopic origin of entropy production and irreversibility. In classical stochastic thermodynamics and in many formulations of nonequilibrium quantum thermodynamics, entropy production is introduced phenomenologically through entropy-balance relations or through relative entropy with respect to a reference equilibrium state [40–44]. Such approaches typically rely on stationary Gibbs states, asymptotic equilibrium distributions, or steady-state assumptions. A formulation of entropy production that remains valid throughout the transient nonequilibrium evolution of an open quantum system, without reference to an equilibrium state, remains highly desirable.

Thermodynamics is fundamentally concerned with energy exchange between a system and its surroundings. Since entropy is fundamentally defined in terms of the probability distribution over accessible microstates, the statistics associated with energy measurements naturally occupy a distinguished role in thermodynamics. This observation suggests that the entropy associated with energy measurement statistics, namely the energy entropy, should play a fundamental role in nonequilibrium thermodynamics [36, 38]. Nevertheless, its role as the appropriate thermodynamic entropy governing coherent nonequilibrium quantum systems has remained largely unexplored.

In this work, we develop a general framework for nonequilibrium quantum thermodynamics by integrating the resource theory of quantum coherence with thermodynamic principles. We define the thermodynamic entropy of a coherent quantum system evolving away from thermal equilibrium as the energy entropy associated with the probability distribution of energy measurements and introduce a dynamical temperature that extends the notion of equilibrium temperature to nonequilibrium states. Within this framework, we obtain an exact entropy-balance relation for coherent nonequilibrium quantum systems,

$$\frac{d}{dt}S(\rho(t)) = \frac{1}{T(t)}\frac{dQ(t)}{dt} - \frac{d}{dt}S\left(\rho(t)\|\rho_{\varepsilon}(t)\right), \tag{1}$$

which explicitly separates the entropy flux associated with heat exchange from the coherence-loss contribution to entropy production. This entropy-balance relation provides an operational connection between quantum coherence, heat flow, and total entropy change.

Unlike conventional approaches based on equilibrium reference states or stationary Gibbs states, the present formulation is expressed entirely in terms of instantaneous dynamical quantities and therefore applies throughout the transient nonequilibrium evolution of an open quantum system, long before thermal equilibrium is reached or a steady state is established. Because the coherence contribution enters explicitly into the entropy-balance relation, the framework suggests that thermodynamic entropy generation and heat dissipation may, in principle, be controlled through coherence engineering and coherent manipulation of the system dynamics.

To demonstrate the generality of the formalism, we apply our thermodynamic framework to an exactly solvable model of an open quantum system consisting of a single bosonic mode coupled to a bosonic reservoir at finite temperature. Using an initial coherent state of the system, we investigate the nonequilibrium dynamics of thermodynamic quantities such as internal energy, thermodynamic entropy, dynamical temperature, work, heat, and free energy, and demonstrate how equilibrium thermodynamics is dynamically recovered in the long-time limit. The first and second laws of thermodynamics are shown to remain valid throughout the transient nonequilibrium evolution. We also show that the generalized entropy balance relation remains valid throughout the transient evolution. Equilibrium thermodynamics emerges naturally in the long-time limit as coherence disappears and the system approaches thermal equilibrium.

The remainder of the paper is organized as follows. In Section 2, we develop the general framework of nonequilibrium quantum thermodynamics and obtain the nonequilibrium entropy-balance equation. In Section 3, we introduce the exactly solvable open quantum model and analyze the dynamical evolution of thermodynamic quantities. Finally, Section 4 summarizes our conclusions and discusses future directions for nonequilibrium quantum thermodynamics and quantum thermal technologies.

# 2 General Framework of Nonequilibrium Quantum Thermodynamics

## 2.1 Open quantum systems and the first law of thermodynamics

An open quantum system [45, 46] coupled to a thermal reservoir is described by the total Hamiltonian,

$$H_{tot}(t) = H(t) + H_R + H_I, \tag{2}$$

where $H(t)$ and $H_R$ are system and reservoir Hamiltonians, and $H_I$ describes the interaction between them. The time evolution of the total density operator of the system and reservoir is governed by the Liouville–von Neumann equation, $i\hbar\dot{\rho}_{tot}(t) = [H_{tot}(t), \rho_{tot}(t)]$. The reservoir is initially considered in thermal equilibrium, while the system is prepared in an arbitrary quantum state, such that the initial total density operator is factorized. The evolution of the open quantum system is then entirely captured by its reduced density operator $\rho(t)$, obtained by eliminating the environmental degrees of freedom from the total system-reservoir density matrix $\rho_{tot}(t)$ through a partial trace, $\rho(t) = \mathrm{tr}_R\{\rho_{tot}(t)\}$. The resulting dynamics of the open system can be expressed using a quantum master equation of the form

$$\frac{d\rho(t)}{dt} = \mathcal{L}(t)\rho(t), \tag{3}$$

where $\mathcal{L}(t)$ is the Liouville–von Neumann super-operator, which contains a Hamiltonian part and a dissipative part (accounting for the influence of the reservoir).

The average energy of the system at an arbitrary time, namely the nonequilibrium internal energy, is given by

$$U(t) = \mathrm{tr}\{H(t)\rho(t)\}. \tag{4}$$

The first law of nonequilibrium quantum thermodynamics follows directly from the time derivative of the internal energy as

$$\dot{U}(t) = \dot{W}(t) + \dot{Q}(t), \tag{5}$$

where the first term on the right of (5) is the rate of change of work given by

$$\dot{W}(t) = \mathrm{tr}\{\dot{H}(t)\rho(t)\}, \tag{6}$$

and the second term on the right of (5) is the rate of heat exchange between the system and the reservoir

$$\dot{Q}(t) = \mathrm{tr}\{H(t)\dot{\rho}(t)\} = \mathrm{tr}\{H(t)\mathcal{L}(t)\rho(t)\}. \tag{7}$$

Following the standard Alicki formulation of quantum thermodynamics [10, 47], the nonequilibrium change in internal energy can be decomposed into two physically distinct contributions. The work term arises from changes in the instantaneous energy spectrum induced by variations in the system Hamiltonian, whereas the heat contribution originates from changes in the occupation probabilities of the energy levels. In quantum systems, the energy spectrum serves as the analogue of volume in classical thermodynamics. Accordingly, variations in the energy levels represent the quantum analogue of mechanical work, whereas changes in the populations of the energy levels are associated with heat exchange and thermodynamic entropy. If the system evolves from time $t = t_0$ to $t = \tau$, the total change in internal energy $\Delta U(\tau) = U(\tau) - U(t_0)$ will be balanced by the work $\Delta W(\tau) = \int_{t_0}^{\tau} \mathrm{tr}\{\dot{H}(t)\rho(t)\}dt$, and heat contribution $\Delta Q(\tau) = \int_{t_0}^{\tau} \mathrm{tr}\{H(t)\mathcal{L}(t)\rho(t)\}dt$.

## 2.2 Thermodynamic entropy and temperature in coherent quantum systems beyond equilibrium

At the heart of nonequilibrium quantum thermodynamics lies the definition of thermodynamic entropy $\mathcal{S}_{th}(t)$ and its relation to heat exchange. In many studies, nonequilibrium entropy is identified with the von Neumann entropy $S(\rho(t))$, multiplied by the Boltzmann constant $k$, and this identification is routinely employed in analyses of thermodynamic laws and in modeling quantum thermal devices. While such an identification is justified in thermal equilibrium, where the system is described by the Gibbs state

$$\rho = \frac{e^{-\beta H}}{Z}, \quad Z = \mathrm{tr}\{e^{-\beta H}\}, \tag{8}$$

with $\beta = 1/kT$ and the density matrix is diagonal in the energy eigenbasis, the situation changes fundamentally away from equilibrium. In equilibrium, the von Neumann entropy [35]

$$S(\rho) = -\mathrm{tr}\{\rho \ln \rho\} \tag{9}$$

coincides with the thermodynamic entropy $\mathcal{S}_{th}$. However, extending this equivalence to nonequilibrium quantum states is generally unjustified [8, 36–39], particularly in the presence of quantum coherence. In the presence of coherence in the energy eigenbasis, the von Neumann entropy does not, in general, coincide with the thermodynamic entropy governing energy exchange and heat flow.

Now, entropy is fundamentally defined in terms of a probability distribution. For quantum systems, such probabilities arise naturally from measurements of observables. For an observable $A = \sum_n a_n |a_n\rangle\langle a_n|$, a system described by the state $\rho(t)$ yields measurement probabilities $w_n(t) = \mathrm{tr}\,(\rho(t)|a_n\rangle\langle a_n|)$, and the associated entropy is $-\sum_n w_n(t) \ln w_n(t)$. Different observables generally lead to different probability distributions and therefore to different entropies. Thermodynamics is fundamentally a theory of energy exchange between a system and its surroundings. Internal energy, heat, work, and the first law are all formulated in terms of energy. Consequently, among all quantum observables, energy occupies a unique operational role in thermodynamics. The probability distribution associated with energy measurements therefore provides the natural statistical description from which thermodynamic entropy should be constructed.

Among all observables, energy therefore occupies a distinguished role in thermodynamics and is represented by the Hamiltonian $H(t) = \sum_n \varepsilon_n(t)|\varepsilon_n(t)\rangle\langle\varepsilon_n(t)|$ of the system, where $\varepsilon_n(t)$ are the instantaneous energy eigenvalues and $|\varepsilon_n(t)\rangle$ are the instantaneous energy eigenstates. If we express the nonequilibrium density matrix $\rho(t)$ with respect to instantaneous energy eigenbasis $\{|\varepsilon_n(t)\rangle\}$, then the diagonal density matrix $\rho_\varepsilon(t)$ is associated with this energy measurement. The diagonal density matrix $\rho_\varepsilon(t)$ can be obtained by removing all off-diagonal elements of $\rho(t)$ in the energy eigenbasis. The diagonal elements $p_n(t) = \mathrm{tr}\,(\rho(t)|\varepsilon_n(t)\rangle\langle\varepsilon_n(t)|)$ of the density matrix represent the probabilities of occupying various energy eigenstates, and the corresponding entropy $S(\rho_\varepsilon(t)) = -\sum_n p_n(t) \ln p_n(t)$ is the energy entropy [36, 38]. With the above physical motivation, we define the thermodynamic entropy $\mathcal{S}_{th}(t)$ of a coherent quantum system evolving away from thermal equilibrium, as the energy entropy associated with the probability distribution of energy measurements. Equivalently, it is the von Neumann entropy of the diagonal density matrix $\rho_\varepsilon(t)$ in the instantaneous energy eigenbasis given by

$$\mathcal{S}_{th}(t) = S(\rho_\varepsilon(t)) = -\sum_n p_n(t) \ln p_n(t). \tag{10}$$

On the other hand, from an information-theoretic perspective [35], the von Neumann entropy $S(\rho(t)) = -\mathrm{tr}\{\rho(t) \ln \rho(t)\}$ corresponds to the entropy of the most informative observable. While von Neumann entropy coincides with thermodynamic entropy in thermal equilibrium, it does not, in general, represent thermodynamic entropy for nonequilibrium states, particularly in presence of quantum coherence in the energy basis. Whenever the nonequilibrium density matrix possesses coherence in the energy eigenbasis, the von Neumann entropy and the thermodynamic entropy are not the same. Away from equilibrium, identifying the von Neumann entropy with thermodynamic entropy can therefore be misleading. The identification of thermodynamic entropy $\mathcal{S}_{th}(t)$ with energy entropy forms the foundation of the nonequilibrium thermodynamic framework developed below.

Having identified the thermodynamic entropy $\mathcal{S}_{th}(t)$ associated with energy measurement statistics, we introduce the nonequilibrium thermodynamic temperature $T(t)$ which is defined through the relation

$$dQ(t) = T(t)\ d\mathcal{S}_{th}(t), \tag{11}$$

where $d\mathcal{S}_{th}(t) = dQ(t)/T(t)$ is the change in thermodynamic entropy due to system-reservoir heat exchange, $T(t)$ is the instantaneous temperature of the system, and $dQ(t)$ is the heat exchange between the system and the reservoir in nonequilibrium. It is important to note that $d\mathcal{S}_{th}(t)$ does not, within the present framework, account for the total entropy change of the system. The total entropy change is described by the nonequilibrium entropy-balance relation (17), which includes an additional contribution associated with coherence loss. The thermodynamic quantities in nonequilibrium evolve through various quantum states before approaching their thermal equilibrium values. The dynamical temperature of the system at nonequilibrium is obtained directly from equation (11) as

$$T(t) = \frac{dQ(t)}{d\mathcal{S}_{th}(t)} = \frac{dQ(t)}{dt} \Big/ \frac{d\mathcal{S}_{th}(t)}{dt}. \tag{12}$$

The definition (11) provides a natural extension of the concept of equilibrium temperature to nonequilibrium quantum systems. Unlike the equilibrium temperature, which characterizes a stationary Gibbs state, the present temperature is an instantaneous thermodynamic quantity that evolves continuously during the nonequilibrium quantum dynamics. The dynamical temperature $T(t)$ is thermodynamically conjugate to the entropy $\mathcal{S}_{th}(t)$ through Eq. (11), thereby preserving the fundamental conjugate structure of equilibrium thermodynamics. As demonstrated in Section 3, for the exactly solvable open quantum system considered in this work, $T(t)$ evolves continuously during the transient nonequilibrium dynamics and asymptotically approaches the equilibrium temperature in the long-time limit as the system relaxes to thermal equilibrium.

Furthermore, following the Legendre-transform structure underlying equilibrium thermodynamics [1], we generalize the notion of the Helmholtz free energy to nonequilibrium quantum systems by performing a Legendre transformation of the internal energy $U(t)$ with respect to the thermodynamic entropy $\mathcal{S}_{th}(t)$. Since the dynamical temperature $T(t)$ is introduced through the relation $dQ(t) = T(t)\ d\mathcal{S}_{th}(t)$, it plays the role of the thermodynamic variable conjugate to $\mathcal{S}_{th}(t)$. The Legendre transformation therefore replaces the thermodynamic entropy by its conjugate temperature as the thermodynamic variable, thereby yielding the nonequilibrium Helmholtz free energy

$$F(t) = U(t) - T(t)\,\mathcal{S}_{th}(t). \tag{13}$$

The quantity $F(t)$ plays the role of a nonequilibrium thermodynamic potential characterizing the transient nonequilibrium evolution of the system.

### 2.3 Quantum coherence, second law, and thermodynamic irreversibility

We then integrate the resource theory of quantum coherence with quantum thermodynamics. Quantum coherence is a fundamental resource for quantum systems, and its quantification has only recently been formalized within a unified resource-theoretic framework [26, 27]. To quantify quantum coherence, we adopt the widely used relative entropy of coherence. Here, coherence is defined with respect to the instantaneous energy eigenbasis $\{|\varepsilon_n(t)\rangle\}$. For an arbitrary nonequilibrium quantum state $\rho(t)$, the relative entropy of coherence is given by [26]

$$C(t) = C(\rho(t)) = \min_{\sigma(t)\in\mathcal{I}} S(\rho(t)\|\sigma(t)), \tag{14}$$

where $S(\rho(t)\|\sigma(t)) = \mathrm{tr}\{\rho(t)\ln\rho(t) - \rho(t)\ln\sigma(t)\}$ measures the distinguishability of $\rho(t)$ from a reference state $\sigma(t)$. The reference states $\sigma(t) \in \mathcal{I}$ are incoherent states in the energy eigenbasis. The minimum in Eq. (14) is evaluated over the set of incoherent states $\mathcal{I}$, which possess no quantum coherence in the energy eigenbasis. It was shown that for the relative entropy of coherence, the closest incoherent state is $\rho_\varepsilon(t) = \sum_n p_n(t)|\varepsilon_n(t)\rangle\langle\varepsilon_n(t)|$, which is the diagonal state of the density matrix $\rho(t)$ obtained by deleting all off-diagonal elements. Hence it is not necessary to perform the minimization to determine the quantum coherence. The relative entropy of coherence (14) is then given by

$$C(t) = S(\rho(t)\|\rho_\varepsilon(t)) = S(\rho_\varepsilon(t)) - S(\rho(t)), \tag{15}$$

where $S(\rho(t))$ and $S(\rho_\varepsilon(t))$ are the von Neumann entropy of the state $\rho(t)$ and the diagonal state $\rho_\varepsilon(t)$, respectively. Using Eq. (15), the decoherence rate or the rate of loss of coherence is then given by

$$-\frac{d}{dt}C(t) = -\frac{d}{dt}S(\rho_\varepsilon(t)) + \frac{d}{dt}S(\rho(t)), \tag{16}$$

By identifying $S(\rho_\varepsilon(t)) = \mathcal{S}_{th}(t)$ and using Eqs. (12), (15), and (16), we obtain the exact entropy-balance relation for nonequilibrium coherent quantum systems as

$$\frac{d}{dt}S(\rho(t)) = \frac{1}{T(t)}\frac{dQ(t)}{dt} - \frac{d}{dt}S(\rho(t)\|\rho_\varepsilon(t)). \tag{17}$$

Within the present framework, Eq. (17) provides a nonequilibrium Clausius relation for coherent quantum systems, in which the entropy change can be written as

$$dS(\rho(t)) = \frac{dQ(t)}{T(t)} + dS_{\text{prod}}(t), \tag{18}$$

where $dS_{\text{prod}}(t) = -dS(\rho(t)\|\rho_\varepsilon(t))$ represents the coherence-loss contribution to entropy production, while $dQ(t)/T(t)$ describes the entropy change associated with the redistribution of energy populations and the corresponding heat exchange between the system and the reservoir. For irreversible decoherence processes under weak system-reservoir coupling, where the relative entropy of coherence decreases monotonically with time, the entropy production is non-negative ($dS_{\text{prod}}(t) \geq 0$). Then for positive entropy production, Eq. (18) naturally leads to the nonequilibrium *Clausius inequality*

$$dS(\rho(t)) \geq \frac{dQ(t)}{T(t)}. \tag{19}$$

## 2.4 Quantum coherence and nonequilibrium entropy production

Equation (17) can be rewritten in the standard entropy-balance form

$$\Sigma(t) = \Phi_Q(t) + \Phi_C(t), \tag{20}$$

where $\Sigma(t) = dS(\rho(t))/dt$ denotes the total entropy change rate, $\Phi_Q(t)$ represents the entropy flux associated with heat exchange, and $\Phi_C(t)$ accounts for the coherence-loss contribution to entropy production. Within the present framework, the entropy-balance relation (20) provides a decomposition of the total entropy change into contributions associated with heat exchange and coherence loss. The first contribution, $\Phi_Q(t)$, represents the entropy flux associated with heat exchange between the system and the reservoir. Its explicit form is given by

$$\Phi_Q(t) = \frac{1}{T(t)}\frac{dQ(t)}{dt}, \tag{21}$$

which is analogous to the corresponding relation in classical stochastic thermodynamics between heat and entropy flux. The quantity $dQ/dt = \text{tr}\{H(t)\mathcal{L}(t)\rho(t)\}$ is the rate of heat flow or the heat current between the system and reservoir, and $T(t)$ is the dynamical temperature of the system at nonequilibrium. Here, $dQ(t)$ is the infinitesimal heat exchanged by the system with the reservoir, and $dQ(t)/T(t)$ is the corresponding thermodynamic entropy supplied to the system by the reservoir under nonequilibrium conditions. The second contribution $\Phi_C(t)$ in the entropy-balance equation (20) arises from the loss of quantum coherence induced by the environment. This contribution represents the component of entropy production associated with coherence loss, which we determine by quantifying the rate of coherence decay in the system. The explicit form of $\Phi_C(t)$ is given by

$$\Phi_C(t) = -\frac{d}{dt}C(t) = -\frac{d}{dt}S(\rho(t)\|\rho_\varepsilon(t)). \tag{22}$$

The entropy-balance relation (20) thus provides an operational connection between quantum coherence, heat flow, and total entropy change in coherent nonequilibrium quantum systems. We restrict ourselves to situations in which there is no particle exchange between the system and the thermal reservoir. The thermal reservoir induces decoherence and dissipation in the energy eigenbasis. In this situation, the total entropy change can be viewed as having two distinct contributions: (i) an entropy contribution associated with heat exchange with the reservoir and (ii) a coherence-loss contribution to entropy production. It is important to note that the entropy-balance relation (20) is formulated for the transient nonequilibrium dynamics, before the system reaches a steady state or thermal equilibrium. Unlike conventional formulations of entropy production based on equilibrium reference states, stationary Gibbs states, or asymptotic thermal states, the present relation is expressed in terms of instantaneous dynamical quantities characterizing the nonequilibrium state of the system. In particular, the relation does not require any reference to a steady state or thermal equilibrium state and remains valid throughout the transient nonequilibrium evolution of the open quantum system. Moreover, the temperature $T(t)$ is not the temperature of the system at thermal equilibrium, but rather a dynamical temperature that evolves with time. In the long-time limit, when the system reaches thermal equilibrium, both the coherence-loss contribution $\Phi_C(t)$ and the entropy flux $\Phi_Q(t)$ approach zero.

By identifying $S(\rho_\varepsilon(t))$ with the thermodynamic entropy $\mathcal{S}_{th}(t)$, Eq. (16) can be written as

$$\frac{d}{dt}S(\rho(t)) = \frac{d}{dt}\mathcal{S}_{th}(t) - \frac{d}{dt}C(t), \tag{23}$$

which relates the von Neumann entropy $S(\rho(t))$, thermodynamic entropy $\mathcal{S}_{th}(t)$, and coherence $C(t)$ during nonequilibrium evolution. Integrating Eq. (23) over a period (from $t = t_0$ to $t = \tau$) of open system dynamics, we obtain

$$\Delta S(\rho(\tau)) = \Delta\mathcal{S}_{th}(\tau) + \Delta C(\tau), \tag{24}$$

where $\Delta S(\rho(\tau)) = S(\rho(\tau)) - S(\rho(t_0))$, $\Delta\mathcal{S}_{th}(\tau) = \mathcal{S}_{th}(\tau) - \mathcal{S}_{th}(t_0)$, and $\Delta C(\tau) = C(t_0) - C(\tau)$ denotes the coherence loss during the nonequilibrium irreversible dynamics of the open system. Equation (24) reveals that entropy generation in coherent nonequilibrium quantum systems has two physically distinct origins. The first contribution, $\Delta\mathcal{S}_{th}(\tau)$, is associated with the thermodynamic entropy transferred through heat exchange between the system and the reservoir. The second contribution, $\Delta C(\tau)$, arises from the loss of quantum coherence and represents entropy generation due to decoherence.

As follows from Eq. (24), the change in coherence enters explicitly into the entropy-balance relation. This indicates that thermodynamic entropy generation can, in principle, be influenced through coherent manipulation of the system. Since thermodynamic entropy is directly associated with heat dissipation, coherence engineering may offer a pathway for regulating thermodynamic losses in quantum thermal devices. In the following section, we demonstrate the nonequilibrium entropy-balance relation and apply the general framework developed above to an exactly solvable model of an open quantum system.

# 3 Quantum Thermodynamics in Nonequilibrium for an Open Quantum System

Next, we apply our nonequilibrium quantum thermodynamics formalism to an open quantum system. We consider an open quantum system comprising a single bosonic mode coupled to a thermal reservoir, described by the Fano-Anderson Hamiltonian [48, 49],

$$\mathscr{H} = H + H_R + H_I = \hbar\omega_0 a^\dagger a + \sum_k \hbar\omega_k b_k^\dagger b_k + \sum_k \hbar \left( V_k a^\dagger b_k + V_k^* b_k^\dagger a \right), \tag{25}$$

where the first term denotes the Hamiltonian of the single-mode bosonic system with frequency $\omega_0$. Here, $a^\dagger$ and $a$ are the creation and annihilation operators of the single bosonic mode. The reservoir is modeled as a collection of bosonic modes with frequencies $\omega_k$, where $b_k^\dagger$ and $b_k$ denote the corresponding creation and annihilation operators, respectively. The parameter $V_k$ represents the coupling amplitude between the system and $k$-th mode of the reservoir. The model Hamiltonian (25) is widely employed in atomic, photonic, and in many-body physics [50–55].

The total system plus reservoir density operator evolves under the Hamiltonian (25) as $\rho_{tot}(t) = e^{-i\mathscr{H}(t-t_0)/\hbar}\rho_{tot}(t_0)e^{i\mathscr{H}(t-t_0)/\hbar}$. We take the initial state of the total system, $\rho_{tot}(t_0) = \rho(t_0) \otimes \rho_R(t_0)$, where $\rho(t_0)$ is the initial state of the system and $\rho_R(t_0) = \exp(-\beta H_R)/\mathrm{tr}\{\exp(-\beta H_R)\}$ is the initial state of the reservoir at thermal equilibrium with inverse temperature $\beta = 1/kT_0$. The nonequilibrium dynamics of the open quantum system can then be fully determined by the reduced density matrix $\rho(t) = \mathrm{tr}_R\{\rho_{tot}(t)\}$. By tracing out the environmental degrees of freedom, an exact master equation governing the system's dynamics can be obtained [56]

$$\begin{aligned} \frac{d}{dt}\rho(t) &= \mathcal{L}(t)\rho(t) \\ &= -\frac{i}{\hbar}\left[H(t), \rho(t)\right] + \gamma(t)\left[2a\rho(t)a^\dagger - a^\dagger a\rho(t) - \rho(t)a^\dagger a\right] \\ &\quad + \widetilde{\gamma}(t)\left[a\rho(t)a^\dagger + a^\dagger\rho(t)a - a^\dagger a\rho(t) - \rho(t)aa^\dagger\right], \end{aligned} \tag{26}$$

where the first term on the right corresponds to a unitary evolution of the system, while the second and third terms describe the nonunitary evolution of the reduced density matrix arising from the system-reservoir interaction, accounting for dissipation and fluctuations, respectively. The nonequilibrium internal energy of the system can be estimated using the renormalized Hamiltonian

$$H(t) = \hbar\omega(t)a^{\dagger}a, \tag{27}$$

which is valid even under strong system-reservoir interaction [57]. The renormalized frequency $\omega(t)$, dissipation coefficient $\gamma(t)$, and the fluctuation coefficient $\widetilde{\gamma}(t)$ in the above quantum master equation (26) are given by $\omega(t) = -\mathrm{Im}\,[\dot{u}(t,t_0)/u(t,t_0)]$, $\gamma(t) = -\mathrm{Re}\,[\dot{u}(t,t_0)/u(t,t_0)]$, and $\widetilde{\gamma}(t) = \dot{v}(t,t) + 2v(t,t)\gamma(t)$. Above time-dependent coefficients of the master equation are completely determined by the functions $u(t,t_0)$ and $v(\tau,t)$, which obey the following integro-differential equations [56]

$$\frac{d}{dt}u(t,t_0) + i\omega_0 u(t,t_0) + \int_{t_0}^{t} d\tau g(t,\tau)u(\tau,t_0) = 0, \tag{28}$$

$$v(\tau,t) = \int_{t_0}^{\tau} d\tau_1 \int_{t_0}^{t} d\tau_2\; u(\tau,\tau_1)\widetilde{g}(\tau_1,\tau_2)u^{*}(t,\tau_2). \tag{29}$$

The integral kernels in Eqs. (28) and (29) are fully determined by the reservoir spectral density $J(\omega)$ through the relations

$$g(t,\tau) = \int_0^{\infty} d\omega J(\omega)e^{-i\omega(t-\tau)}, \tag{30}$$

$$\widetilde{g}(\tau_1,\tau_2) = \int_0^{\infty} d\omega J(\omega)\bar{n}(\omega,T_0)e^{-i\omega(\tau_1-\tau_2)}. \tag{31}$$

The spectral density is defined by $J(\omega) = \sum_k |V_k|^2\delta(\omega-\omega_k)$ and $\bar{n}(\omega,T_0) = 1/\left(e^{\hbar\omega/kT_0}-1\right)$ corresponds to the Bose-Einstein distribution for the bosonic thermal reservoir. We consider an Ohmic spectral density $J(\omega) = \eta\omega\exp(-\omega/\omega_c)$, where $\eta$ denotes the system-reservoir coupling strength and $\omega_c$ is the cutoff frequency of the reservoir spectrum [58]. For this Ohmic reservoir spectrum, localized bound state emerges when the system-reservoir coupling strength exceeds the critical value $\eta_c = \omega_0/\omega_c$.

We consider two different types of initial states for the single-mode bosonic system, one without having coherence (see Supplementary Information), and another one having quantum coherence. First, we investigate our formalism of nonequilibrium quantum thermodynamics in presence of coherence when the thermodynamic entropy $\mathcal{S}_{th}(t)$ is not equal to von Neumann entropy $S(\rho(t))$. We consider the system be prepared initially in a coherent state given by

$$|\alpha_0\rangle = \exp\left(-\frac{1}{2}|\alpha_0|^2\right)\sum_{n=0}^{\infty}\frac{\alpha_0^n}{\sqrt{n!}}|n\rangle, \tag{32}$$

and the reservoir is initially in thermal equilibrium state with an initial temperature $T_0$. Because of the interaction, the system and the reservoir undergo a nonequilibrium process. By solving the quantum master equation (26), one obtains the time evolved reduced density matrix elements of the system [54]

$$\begin{aligned}\langle m|\rho(t)|n\rangle &= e^{-A(t)|\alpha_0|^2}\frac{[\alpha(t)]^m[\alpha^*(t)]^n}{[1+v(t,t)]^{m+n+1}} \\ &\quad\times\sum_{k=0}^{\min\{m,n\}}\frac{\sqrt{m!\;n!}}{(m-k)!\;(n-k)!\;k!}\left[\frac{v(t,t)}{A(t)|\alpha_0|^2}\right]^k, \end{aligned}\tag{33}$$

where $A\left(t\right)=\frac{|u(t,t_0)|^2}{1+v(t,t)}$ and $\alpha(t)=u(t,t_0)\alpha_0$. In this case, the density matrix $\rho(t)$ is not diagonal in the energy eigenbasis. Hence, the von Neumann entropy and thermodynamic entropy are not the same.

To determine the thermodynamic entropy, we first note that the diagonal elements of the density matrix are given by

$$\langle n|\rho(t)|n\rangle = e^{-A(t)|\alpha_0|^2}\frac{|\alpha(t)|^{2n}}{[1+v(t,t)]^{2n+1}}\sum_{k=0}^{n}\frac{n!}{[(n-k)!]^2\;k!}\left[\frac{v(t,t)}{A(t)|\alpha_0|^2}\right]^k. \tag{34}$$

These diagonal elements of the density matrix define the probabilities $p_n^{\alpha_0}(t)$ of obtaining the corresponding energy eigenvalues in a measurement of the Hamiltonian $H(t)$ given in Eq. (27). For the open quantum system considered here, the instantaneous energy eigenstates $\{|\varepsilon_n(t)\rangle\}$ of the Hamiltonian $H(t)$ coincide with the Fock basis states $\{|n\rangle\}$, whereas the corresponding instantaneous energy eigenvalues $\varepsilon_n(t)=n\hbar\omega(t)$ are explicitly time dependent. Consequently, $p_n^{\alpha_0}(t)=\langle n|\rho(t)|n\rangle$ represents the probability of finding the system in the energy eigenstate $|n\rangle$ at time $t$. The entropy associated with this energy-measurement probability distribution is the energy entropy. Following the analysis of entropy production presented in the previous section, the thermodynamic entropy of the nonequilibrium open quantum system is given by the energy entropy,

$$\mathcal{S}_{th}(t) = -\sum_{n=0}^{\infty} p_n^{\alpha_0}(t)\ln p_n^{\alpha_0}(t), \tag{35}$$

which can, in principle, be experimentally determined from the probability distribution $p_n^{\alpha_0}(t)$.

Next, we evaluate the nonequilibrium von Neumann entropy $S(\rho(t))$ of the open quantum system coupled to the thermal reservoir. The single-mode bosonic system considered here is a continuous-variable quantum system [59, 60]. The system is initially prepared in a coherent state (32), belonging to the class of Gaussian states, which are characterized by a Gaussian Wigner function [61, 62]. For a single-mode bosonic

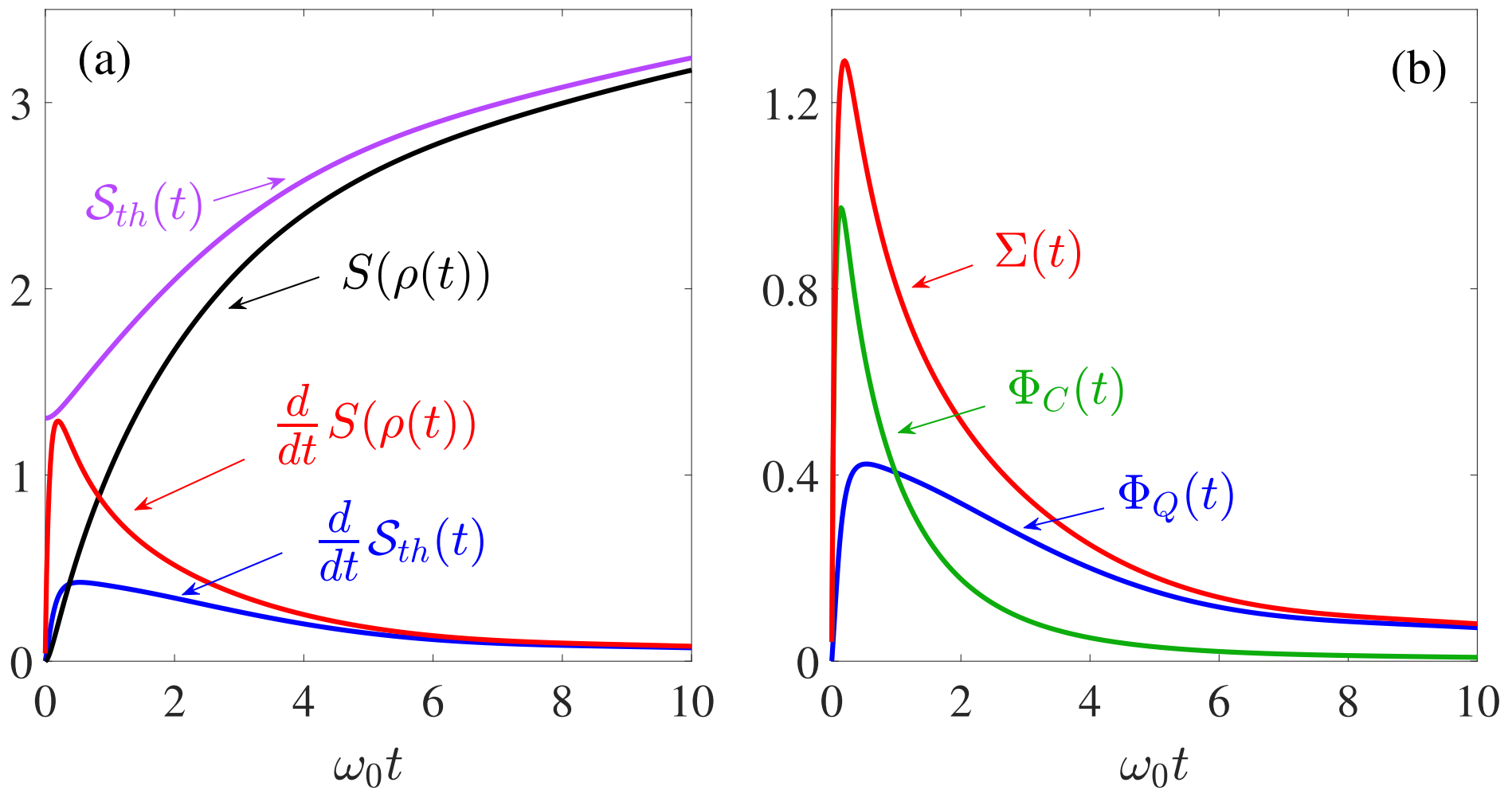

**Fig. 1** Entropy balance far from equilibrium: (a) thermodynamic entropy $\mathcal{S}_{th}(t)$ (purple), von Neumann entropy $S(\rho(t))$ (black), and their derivatives $\dot{\mathcal{S}}_{th}(t)$ (blue) and $\dot{S}(\rho(t))$ (red) are shown as a function of time (b) total entropy rate $\Sigma(t)$ (red), entropy production due to coherence loss $\Phi_C(t)$ (green), and entropy flux rate due to heat exchange $\Phi_Q(t)$ (blue) are shown as a function of time much before the system reaches equilibrium. The system is initially prepared in a coherent state $|\alpha_0\rangle$, with coherent state amplitude $\alpha_0 = 1$. The initial temperature of the reservoir is fixed at $kT_0 = 20\hbar\omega_0$. The cutoff frequency $\omega_c = 10\omega_0$, and the system-reservoir coupling strength $\eta = 0.1\eta_c$

system prepared in a Gaussian state, the von Neumann entropy can be directly evaluated from the covariance matrix $\boldsymbol{V}(t)$. The covariance matrix is fully determined by the first and second moments of the quadrature operators. The quadrature vector $\boldsymbol{\xi} = \{\xi_1, \xi_2\}$ comprises the canonical quadratures $\xi_1 = (a + a^\dagger)$ and $\xi_2 = -i(a - a^\dagger)$. The elements of the covariance matrix are given by

$$V_{ii}(t) = \langle \xi_i^2 \rangle - \langle \xi_i \rangle^2, \text{ and } V_{ij}(t) = \frac{1}{2}\langle \xi_i \xi_j + \xi_j \xi_i \rangle - \langle \xi_i \rangle \langle \xi_j \rangle, \tag{36}$$

where the averages are taken with respect to the nonequilibrium density matrix $\rho(t)$. Because of the structure of the Hamiltonian (25), the quantum state retains its Gaussian character during the entire time evolution. Consequently, the von Neumann entropy of the nonequilibrium Gaussian state $\rho(t)$ is given by [61, 63]

$$S(\rho(t)) = \frac{\nu(t)+1}{2} \ln \frac{\nu(t)+1}{2} - \frac{\nu(t)-1}{2} \ln \frac{\nu(t)-1}{2}, \tag{37}$$

where $\nu(t) = \sqrt{\det \boldsymbol{V}(t)}$. To determine the covariance matrix elements and the von Neumann entropy under nonequilibrium conditions, we employ the Heisenberg equations of motion to evaluate the time evolution of the quadrature moments [64]. Details of the calculation are provided in the Supplementary Information.

Before we go to nonequilibrium thermodynamic quantities such as heat, work, temperature, and free energy, we first investigate the entropy-balance relation (20) for this open system model. The system is initially prepared in a coherent state $|\alpha_0\rangle$, and the initial reservoir state is taken at thermal equilibrium with initial temperature $kT_0 = 20\hbar\omega_0$. We consider weak system-reservoir coupling ($\eta = 0.1\eta_c$). Using the reduced density matrix $\rho(t)$ (expressed in energy basis) given by Eq. (33), we compute the time-dependent von Neumann entropy $S(\rho(t))$ using Eq. (37). The nonequilibrium thermodynamic entropy $\mathcal{S}_{th}(t)$, defined in Eq. (35), is evaluated through the diagonal density matrix $\rho_\varepsilon(t)$ with matrix elements given by Eq. (34). The diagonal elements of the density matrix represent the probabilities $p_n^{\alpha_0}(t)$ associated to energy measurement. The entropy production due to coherence loss $\Phi_C(t)$ is then calculated using Eq. (22). The entropy flux $\Phi_Q(t)$ due to heat exchange between the system and reservoir is calculated using Eq. (21). Figure 1a provides direct numerical evidence that the von Neumann entropy $S(\rho(t))$ does not match with the thermodynamic entropy $\mathcal{S}_{th}(t)$ during coherent nonequilibrium evolution.

Fig. 1a shows that both $\mathcal{S}_{th}(t)$ and $S(\rho(t))$ increase monotonically as the system exchanges energy and information with the reservoir through a nonequilibrium process. In nonequilibrium, the value of thermodynamic entropy $\mathcal{S}_{th}(t)$ at any instant of time is higher than the von Neumann entropy, *i.e.* $\mathcal{S}_{th}(t) \geq S(\rho(t))$. The equality, $\mathcal{S}_{th}(t) = S(\rho(t))$, holds in the long-time limit, when the system reaches thermal equilibrium. We note that although $S(\rho(t)) \leq \mathcal{S}_{th}(t)$ in general, the entropy rate $\Sigma(t) = dS(\rho(t))/dt$ is always higher than the thermodynamic entropy rate $d\mathcal{S}_{th}(t)/dt$. Fig. 1b shows that initially for a short interval of time, the entropy production due to coherence loss $\Phi_C(t)$ dominates over the entropy flux $\Phi_Q(t)$ due to heat exchange. After a while, the entropy flux $\Phi_Q(t)$ due to heat exchange dominates over the entropy production due to decoherence $\Phi_C(t)$ quantified by the rate of coherence loss. In nonequilibrium, there is a competition of these two processes of decoherence and heat exchange. But, the entropy balance $\Sigma(t) = \Phi_Q(t) + \Phi_C(t)$ is always satisfied at any given moment in nonequilibrium. From Fig. 1b, we see that the entropy production $\Phi_C(t)$ by the system in nonequilibrium

$$\Phi_C(t) = -\frac{d}{dt}C(t) = -\frac{d}{dt}S(\rho(t)\|\rho_\varepsilon(t)) \geq 0 \tag{38}$$

under weak system-reservoir coupling. In the long-time limit, the system reaches thermal equilibrium, the entropy production $\Phi_C(t)$ and entropy flux $\Phi_Q(t)$ both approach zero.

Using Eq. (27), the average energy (4) of the bosonic open system at an arbitrary time, namely the nonequilibrium internal energy, is given by

$$U(t) = \mathrm{tr}\,[H(t)\rho(t)] = \hbar\,\omega(t)\;n(t), \tag{39}$$

where the instantaneous average occupation $n(t) = \mathrm{tr}[a^\dagger a\rho(t)] = |u(t,t_0)|^2|\alpha_0|^2 + v(t,t)$. Using the first law of nonequilibrium quantum thermodynamics (5), the rate

of change of work (6) for this open system is given by

$$\dot{W}(t) = \mathrm{tr}\{\dot{H}(t)\rho(t)\} = \hbar\, \dot{\omega}(t)\, n(t), \tag{40}$$

and the rate of heat flow (7) between the system and the reservoir is given by

$$\dot{Q}(t) = \mathrm{tr}\{H(t)\dot{\rho}(t)\} = \hbar\omega(t)\Big\{\widetilde{\gamma}(t) - 2\gamma(t)n(t)\Big\}. \tag{41}$$

Although the present model contains no explicit external driving field, the exact reduced dynamics generate a time-dependent renormalized system frequency $\omega(t)$ through the interaction with the reservoir. This transient frequency renormalization (Lamb shift) induces a corresponding transient work contribution through equation (40). From these results, one can determine the dynamical temperature $T(t)$ of the system defined by Eq. (12) as $T(t) = \dot{Q}(t)/\dot{\mathcal{S}}_{th}(t)$. During nonequilibrium evolution, the dynamical temperature $T(t)$ of the system generally differs from the initial reservoir temperature $T_0$. We set the Boltzmann constant to unity ($k = 1$), so that temperature is measured on an energy scale and entropy is dimensionless. Accordingly, the dynamical temperature $T(t)$ carries units of energy and is expressed in units of $\hbar\omega_0$. To recover the temperature in Kelvin, we need to multiply $T(t)$ by $\hbar\omega_0$, and then have to divide it by $k$, as the Boltzmann $k$ sets the energy scale corresponding to one Kelvin.

We now apply our nonequilibrium quantum thermodynamic framework to evaluate key thermodynamic quantities. The open quantum system is prepared initially in a coherent state $|\alpha_0\rangle$, with $\alpha_0 = 1$. Using the reduced density matrix $\rho(t)$ given by Eq. (33), we compute the time-dependent internal energy $U(t)$ via Eq. (39). The nonequilibrium thermodynamic entropy $\mathcal{S}_{th}(t)$, defined in Eq. (35), is evaluated through the probability distribution of energy states given by Eq. (34). Combining thermodynamic entropy $\mathcal{S}_{th}(t)$ in Eq. (35) with the heat current given by Eq. (41), we determine the dynamical temperature $T(t)$ from Eq. (12). The nonequilibrium free energy $F(t)$ is then calculated using Eq. (13). The resulting dynamics of $U(t)$, $\mathcal{S}_{th}(t)$, $T(t)$, and $F(t)$ are shown in Figs. 2a-2d, respectively, for weak system-reservoir coupling and different initial reservoir temperatures $T_0$. As shown in Figs. 2a and 2b, both the internal energy and the thermodynamic entropy increase monotonically as the system exchanges energy and information with the reservoir through a nonequilibrium process, eventually saturating at values determined by $T_0$. The dynamical temperature $T(t)$ rises gradually (Fig. 2c) and asymptotically approaches the reservoir temperature $T_0$, signalling thermal equilibration. These results provide a dynamical picture of thermalization in the weak-coupling limit, demonstrating how equilibrium thermodynamics emerges from the underlying quantum dynamics. Consistent with the second law of thermodynamics, the thermodynamic entropy $\mathcal{S}_{th}(t)$ increases throughout the nonequilibrium evolution (Fig. 2b). Correspondingly, the free energy $F(t)$ decreases monotonically (Fig. 2d), in accordance with the constraints on nonequilibrium state transformations [65]. The simultaneous increase of internal energy and decrease of free energy implies a net heat flow into the system, leading to the observed rise in the dynamical temperature $T(t)$. In the long-time limit, the system reaches thermal equilibrium, characterized by maximal entropy and minimal free energy.

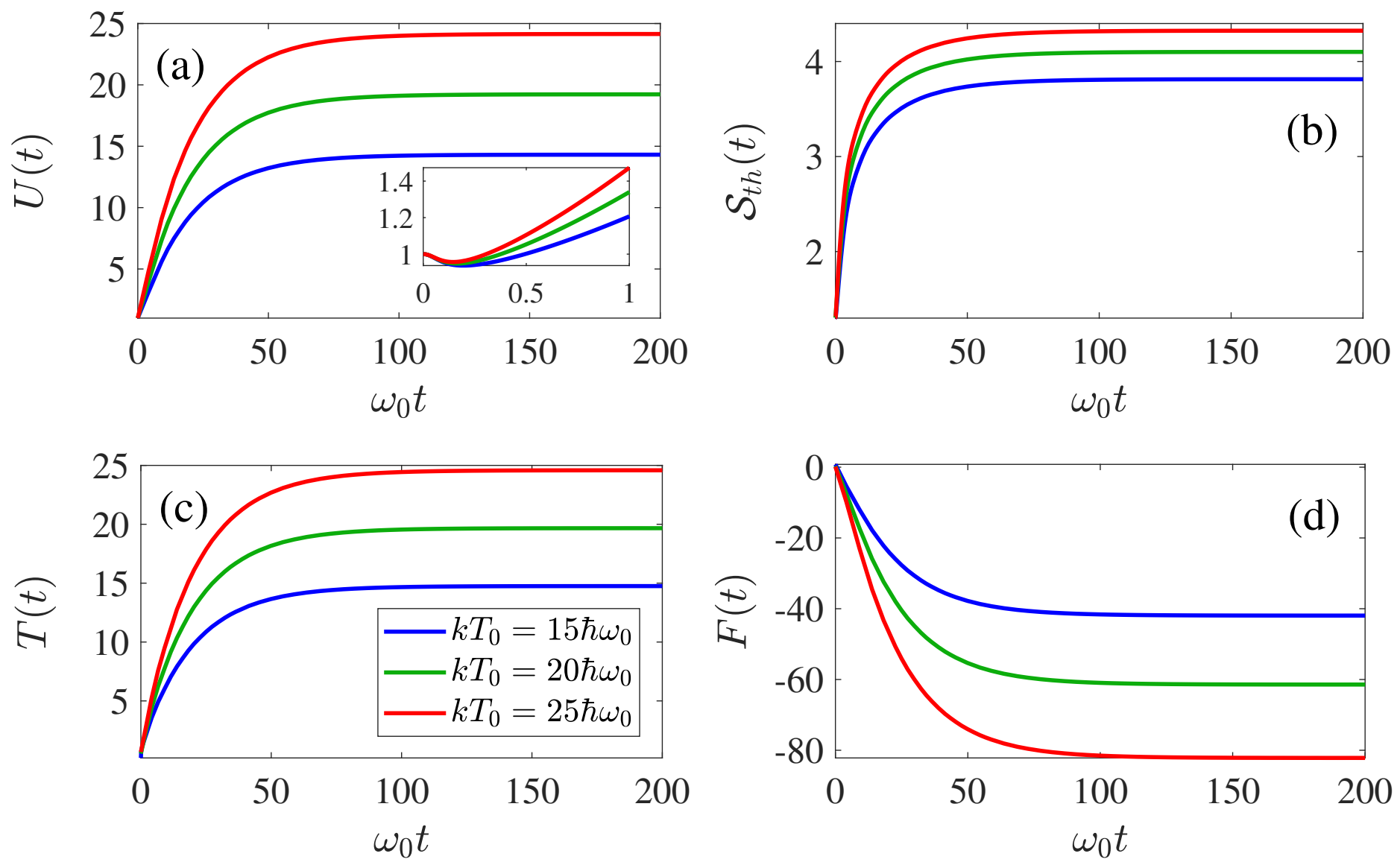

**Fig. 2** Quantum thermodynamic quantities under nonequilibrium dynamics: Time evolution of (a) the internal energy $U(t)$, (b) the thermodynamic entropy $\mathcal{S}_{th}(t)$, (c) the dynamical temperature $T(t)$, and (d) the free energy $F(t)$, evaluated from Eqs. (35)–(13). The system is initially prepared in a coherent state $|\alpha_0\rangle$, with $\alpha_0 = 1$. In each panel, the three curves correspond to different initial temperatures of the thermal reservoir: $kT_0 = 15\hbar\omega_0$ (blue), $20\hbar\omega_0$ (green), and $25\hbar\omega_0$ (red). The system-reservoir coupling strength is fixed at $\eta = 0.1\eta_c$, with a reservoir cutoff frequency $\omega_c = 10\omega_0$.

Next, we examine the first law of thermodynamics (5), which dictates that the rate of change of the system's internal energy is balanced by the work rate (6) and the heat current (7). For the coherent open system, the rate of change of internal energy $\dot{U}(t)$ is evaluated from Eq. (39), while the work rate $\dot{W}(t)$ and the heat flow rate $\dot{Q}(t)$ are obtained from Eqs. (40) and (41), respectively. Figures 3a-3c show the time evolution of $\dot{U}(t)$, $\dot{W}(t)$, and $\dot{Q}(t)$ for different initial reservoir temperatures $T_0$, when the system is far from equilibrium. At short times, the internal energy decreases transiently, resulting in negative values of $\dot{U}(t)$. This behavior is consistent with the short-time dynamics of $U(t)$ shown in the inset of Fig. 2a. The negative work rate $\dot{W}(t)$ in Fig. 3b indicates that quantum work is done on the system by the reservoir during its nonequilibrium time evolution. We find that variations in the initial reservoir temperature have a negligible effect on the work rate. At later times, the internal energy increases due to reservoir-induced heating, accompanied by a positive heat current $\dot{Q}(t)$, signifying net heat flow from the reservoir into the system (see Fig. 3c). The heat current reflects the combined influence of dissipative and fluctuation dynamics, as encoded in Eq. (41). In the long-time limit, both $\dot{W}(t)$ and $\dot{Q}(t)$ approach zero as the system relaxes to thermal equilibrium. Finally, we perform the nonequilibrium thermodynamic analysis of the open quantum system in the absence of quantum coherence. We assume that the system is initially prepared in a Fock state $|n_0\rangle$ and evolves under the system-reservoir interaction. In this case, the time-evolved state remains diagonal in the energy eigenbasis. The corresponding results are presented in the Supplementary Information (see Supplementary Fig. S1 and Supplementary Fig. S2).

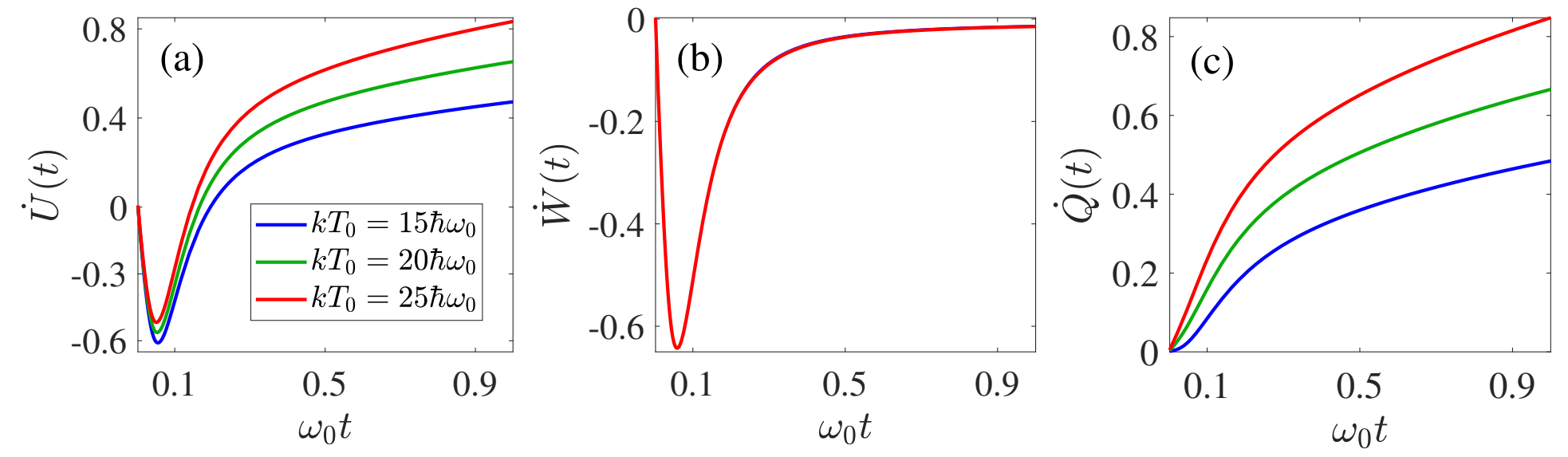

**Fig. 3** Nonequilibrium dynamics of internal energy, work, and heat exchange rates (in unit of $\hbar\omega_0^2$): Time evolution of (a) the rate of change of internal energy $\dot{U}(t)$, (b) the work rate $\dot{W}(t)$, and (c) the heat flow rate $\dot{Q}(t)$, evaluated from Eqs. (39), (40), and (41), respectively. The system is initially prepared in a coherent state $|\alpha_0\rangle$ with $\alpha_0 = 1$. We consider three different initial temperatures of the thermal reservoir: $kT_0 = 15\hbar\omega_0$ (blue), $20\hbar\omega_0$ (green), and $25\hbar\omega_0$ (red). The work rate dynamics exhibits negligible sensitivity to reservoir temperature $T_0$. The system-reservoir coupling strength is fixed at $\eta = 0.1\eta_c$, with a reservoir cutoff frequency $\omega_c = 10\omega_0$.

# 4 Conclusions

We have developed a framework for nonequilibrium quantum thermodynamics by integrating the quantum resource theory of coherence with thermodynamic laws. Central to this formulation is an exact entropy-balance equation and the resulting nonequilibrium Clausius relation, which apply throughout the transient nonequilibrium evolution and explicitly separate the entropy flux due to heat exchange from the coherence-loss contribution to entropy production. This decomposition provides a transparent description of the role of coherence in quantum thermodynamics.

A key concept of our analysis is the identification of the appropriate thermodynamic entropy in nonequilibrium quantum processes. We find that, in the presence of coherence, the von Neumann entropy does not generally represent thermodynamic entropy. Instead, the entropy associated with energy measurements, namely the energy entropy defined by the diagonal populations in the energy eigenbasis, governs the thermodynamic description of heat exchange. This provides a physically motivated framework for addressing a long-standing ambiguity in the formulation of quantum thermodynamics beyond equilibrium.

Applying the framework to an exactly solvable open quantum system, we demonstrate that the first and second laws of thermodynamics are found to be satisfied throughout the transient nonequilibrium evolution, with equilibrium thermodynamics emerging dynamically in the weak-coupling limit. In contrast to conventional approaches based on equilibrium reference states or stationary Gibbs ensembles, the present formulation is expressed in terms of instantaneous dynamical quantities and can therefore be applied far from equilibrium. Our results provide an operational connection between quantum coherence, heat flow, and total entropy change, offering a consistent framework for describing coherent nonequilibrium thermodynamic processes.

The present framework is applicable to a wide class of driven and dissipative quantum systems and has potential implications for quantum thermal machines. Because quantum coherence enters explicitly into the entropy-balance relation, the formulation suggests that thermodynamic entropy changes and heat dissipation may, in principle, be influenced through coherence engineering and coherent manipulation of the system dynamics. This suggests opportunities for designing quantum thermal technologies with improved control over thermodynamic losses. Furthermore, because the entropy-balance equation and the resulting nonequilibrium Clausius relation are formulated in terms of experimentally accessible energy statistics and quantum coherence measures, the framework provides a natural pathway toward experimental investigation in cavity and circuit QED systems, superconducting qubits, trapped ions, and cold-atom platforms.

# 5 Supplementary Material

The Supplementary Material contains the detailed derivation of the von Neumann entropy for Gaussian states using the covariance-matrix formalism, complementary theoretical analysis of nonequilibrium quantum thermodynamics for an initially incoherent Fock state, and additional numerical results illustrating the evolution of thermodynamic quantities, work, heat, entropy, temperature, and free energy. These supplementary results further validate the generality and consistency of the nonequilibrium thermodynamic framework developed in the main text.

**Data Availability**
The authors declare that the data supporting the findings of this study are available within the paper and its supplementary information file.

**Acknowledgements**
M.M.A. was supported by the Centre for Quantum Science and Technology, Chennai Institute of Technology, India.

**Author Contributions**
M.M.A. conceived the original idea and developed the formulation. P.W.C. and M.M.A. performed the calculations and interpreted the results. All authors contributed to the research and the preparation of the manuscript.

**Funding**
P.W.C. received funding support from the National Atomic Research Institute

(NARI), Taiwan. No specific grant number.

**Competing Interests**
The authors declare no competing interests.

# Supplementary Information for "Thermodynamics Beyond Equilibrium in Coherent Quantum Systems"

Md Manirul Ali[1*] and Po-Wen Chen[2]

[1]Centre for Quantum Science and Technology, Chennai Institute of Technology, Sarathy Nagar, Kundrathur, Chennai, 600069, India.
[2]Department of Physics, National Atomic Research Institute, Taoyuan, 325207, Taiwan.

*Corresponding author(s). E-mail(s): manirul@citchennai.net;
Contributing authors: powen@nari.org.tw;

## 1 von Neumann entropy of Gaussian states: Initial state is coherent state $|\alpha_0\rangle$

First, we considered the system be prepared initially in a coherent state given by

$$|\alpha_0\rangle = \exp\left(-\frac{1}{2}|\alpha_0|^2\right)\sum_{n=0}^{\infty}\frac{\alpha_0^n}{\sqrt{n!}}|n\rangle \tag{1}$$

Here we provide the detailed calculation to estimate the von Neumann entropy $S(\rho(t))$ of the open quantum system in nonequilibrium. We considered a single-mode bosonic system, which constitutes a continuous-variable quantum system [1, 2]. The system is initially prepared in a coherent state (1), belonging to the class of Gaussian states [3, 4]. For a single-mode bosonic system prepared in a Gaussian state, the von Neumann entropy can be directly evaluated from the covariance matrix $\boldsymbol{V}(t)$. The covariance matrix is fully determined by the first and second moments of the quadrature operators. The quadrature vector $\boldsymbol{\xi} = \{\xi_1, \xi_2\}$ comprises the canonical quadratures $\xi_1 = (a + a^\dagger)$ and $\xi_2 = -i(a - a^\dagger)$. The covariance matrix elements are

given by [3]

$$V_{ij}(t) = \frac{1}{2}\langle\{\Delta\xi_i, \Delta\xi_j\}\rangle = \frac{1}{2}\langle(\Delta\xi_i\Delta\xi_j + \Delta\xi_j\Delta\xi_i)\rangle, \tag{2}$$

where the averages are taken with respect to the nonequilibrium density matrix $\rho(t)$, and

$$\Delta\xi_i = \xi_i - \langle\xi_i\rangle. \tag{3}$$

Using Eq. (3), the covariance matrix elements can further be simplified as

$$V_{ii}(t) = \langle\xi_i^2\rangle - \langle\xi_i\rangle^2, \text{ and } V_{ij}(t) = \frac{1}{2}\langle\xi_i\xi_j + \xi_j\xi_i\rangle - \langle\xi_i\rangle\langle\xi_j\rangle. \tag{4}$$

Because of the structure of the Hamiltonian (6), the quantum state retains its Gaussian character during the entire time evolution. Consequently, the von Neumann entropy of a Gaussian state $\rho(t)$ is given by [3, 5]

$$S(\rho(t)) = \frac{\nu(t)+1}{2}\ln\frac{\nu(t)+1}{2} - \frac{\nu(t)-1}{2}\ln\frac{\nu(t)-1}{2}, \tag{5}$$

where $\nu(t) = \sqrt{\det \boldsymbol{V}(t)}$. To determine the covariance matrix elements in nonequilibrium, we use Heisenberg approach to evaluate the time evolution of quadrature moments. We consider the open quantum system comprising of a single bosonic mode coupled to a thermal reservoir, described by the Fano-Anderson Hamiltonian [6–11]

$$\mathscr{H} = H + H_R + H_I = \hbar\omega_0 a^\dagger a + \sum_k \hbar\omega_k b_k^\dagger b_k + \sum_k \hbar\Big(V_k a^\dagger b_k + V_k^* b_k^\dagger a\Big). \tag{6}$$

In the Heisenberg picture, the system and reservoir operators evolve as [11]

$$b_k(t) = b_k(t_0)e^{-i\omega_k(t-t_0)} - iV_k^*\int_{t_0}^t d\tau\; a(\tau)\; e^{-i\omega_k(t-\tau)}. \tag{7}$$

$$\dot{a}(t) + i\omega_0 a(t) + \int_{t_0}^t d\tau g(t,\tau)a(\tau) = -i\sum_k V_k b_k(t_0)e^{-i\omega_k(t-t_0)}, \tag{8}$$

where the Heisenberg picture operators $a(t) = e^{i\mathscr{H}(t-t_0)/\hbar}a(t_0)e^{-i\mathscr{H}(t-t_0)/\hbar}$ and $b_k(t) = e^{i\mathscr{H}(t-t_0)/\hbar}b_k(t_0)e^{-i\mathscr{H}(t-t_0)/\hbar}$. For a reservoir with a continuous spectrum, the integral kernel $g(t,\tau) = \sum_k |V_k|^2 e^{-i\omega_k(t-\tau)} = \int_0^\infty d\omega J(\omega)e^{-i\omega(t-\tau)}$, where $J(\omega)$ is the spectral density of the reservoir. The linearity of Eq. (8) ensures the time evolution of $a(t)$ as [12]

$$a(t) = u(t,t_0)a(t_0) + f(t). \tag{9}$$

The time-dependent coefficient $u(t, t_0)$ and the noise operator $f(t)$ satisfy the integro-differential equations

$$\frac{d}{dt}u(t,t_0) = -i\omega_0 u(t,t_0) - \int_{t_0}^{t} d\tau g(t,\tau)u(\tau,t_0), \tag{10}$$

$$\frac{d}{dt}f(t) = -i\omega_0 f(t) - \int_{t_0}^{t} d\tau g(t,\tau)f(\tau) - i\sum_k V_k b_k(t_0)e^{-i\omega_k(t-t_0)}. \tag{11}$$

The coefficient $u(t, t_0)$ is obtained by numerically solving Eq. (10) with the initial condition $u(t_0, t_0) = 1$. In parallel, the noise operator $f(t)$ follows from solving Eq. (11) under the initial condition $f(t_0) = 0$. Nonequilibrium thermal fluctuations are quantified through the correlation function defined below

$$\langle f^\dagger(t)f(t)\rangle = v(t,t) = \int_{t_0}^{t} d\tau_1 \int_{t_0}^{t} d\tau_2 \;\; u(t,\tau_1)\widetilde{g}(\tau_1,\tau_2)u^*(t,\tau_2), \tag{12}$$

where the combined system-reservoir state at the initial time is assumed to be factorized, $\rho_{\text{tot}}(t_0) = \rho(t_0) \otimes \rho_R(t_0)$. The initial state of the reservoir is taken as the thermal equilibrium state $\rho_R(t_0) = \exp(-\beta H_R)/\text{Tr}[\exp(-\beta H_R)]$, where $\beta = 1/kT_0$ is the inverse temperature. The two-time correlation function $\widetilde{g}(\tau_1, \tau_2)$ is given by

$$\widetilde{g}(\tau_1,\tau_2) = \int_0^\infty d\omega J(\omega)\bar{n}(\omega)e^{-i\omega(\tau_1-\tau_2)}. \tag{13}$$

The bosonic reservoir is initially characterized by the Bose-Einstein distribution $\bar{n}(\omega) = 1/(e^{\hbar\omega/kT_0} - 1)$. Using Eqs. (9) to (13), operator averages can be evaluated as

$$\langle a(t)\rangle = u(t,t_0)\langle a(t_0)\rangle, \langle a^\dagger(t)\rangle = u^*(t,t_0)\langle a^\dagger(t_0)\rangle, \langle a(t)a(t)\rangle = (u(t,t_0))^2\langle a(t_0)a(t_0)\rangle, \tag{14}$$

$$\langle a^\dagger(t)a^\dagger(t)\rangle = (u^*(t,t_0))^2\langle a^\dagger(t_0)a^\dagger(t_0)\rangle, \langle a^\dagger(t)a(t)\rangle = |u(t,t_0)|^2\langle a^\dagger(t_0)a(t_0)\rangle + v(t,t). \tag{15}$$

For the above factorized initial system-reservoir state, the noise operators have vanishing mean values, $\langle f(t)\rangle = \langle f^\dagger(t)\rangle = 0$, and $\langle f(t)f(t)\rangle = \langle f^\dagger(t)f^\dagger(t)\rangle = 0$. Using the time-dependent expectation values given in Eqs. (14) and (15), we compute the quadrature moments $\langle\xi_1(t)\rangle$, $\langle\xi_2(t)\rangle$, $\langle\xi_1^2(t)\rangle$, $\langle\xi_2^2(t)\rangle$, $\langle\xi_1(t)\xi_2(t)\rangle$, and $\langle\xi_2(t)\xi_1(t)\rangle$. These moments then determine the covariance matrix elements [13]

$$\begin{aligned} V_{11}(t) &= 1 + 2v(t,t) + 2|u(t,t_0)|^2\,\text{Cov}(a^\dagger(t_0),a(t_0)) \\ &\quad + (u(t,t_0))^2\,\text{Var}(a(t_0)) + (u^*(t,t_0))^2\,\text{Var}(a^\dagger(t_0)), \end{aligned} \tag{16}$$

$$\begin{aligned} V_{22}(t) &= 1 + 2v(t,t) + 2|u(t,t_0)|^2\,\text{Cov}(a^\dagger(t_0),a(t_0)) \\ &\quad - (u(t,t_0))^2\,\text{Var}(a(t_0)) - (u^*(t,t_0))^2\,\text{Var}(a^\dagger(t_0)), \end{aligned} \tag{17}$$

$$V_{12}(t) = i(u^*(t,t_0))^2\,\text{Var}(a^\dagger(t_0)) - i((u(t,t_0))^2\,\text{Var}(a(t_0)). \tag{18}$$

Here, the covariance between two operators is defined as $\text{Cov}(a,b) = \langle ab\rangle - \langle a\rangle\langle b\rangle$, with the variance given by $\text{Var}(a) = \text{Cov}(a,a)$. The covariance matrix is symmetric by construction, implying $V_{12}(t) = V_{21}(t)$. For a given initial system-reservoir state, all time-dependent covariance matrix elements are uniquely determined by the functions $u(t, t_0)$ and $v(t, t)$. To evaluate these functions, we specify the reservoir spectral density and choose an Ohmic form for $J(\omega) = \eta\omega \exp(-\omega/\omega_c)$ [14]. For this Ohmic reservoir spectrum, localized bound state emerges when the system-reservoir coupling strength exceeds the critical value $\eta_c = \omega_0/\omega_c$.

## 2 Nonequilibrium quantum thermodynamics in absence of coherence: Initial Fock state $|n_0\rangle$

We consider again the open quantum system comprising of a single bosonic mode coupled to a thermal reservoir, described by the Fano-Anderson Hamiltonian given by Eq. (6). Now, we consider the system be prepared initially in a Fock state $|n_0\rangle$ and the reservoir is again considered to be in thermal equilibrium initially. The time evolved reduced density matrix of the system is given by [10]

$$\rho(t) = \sum_{n=0}^{\infty} p_n^{n_0}(t)|n\rangle\langle n|, \tag{19}$$

where

$$\begin{aligned} p_n^{n_0}(t) &= \frac{[v(t,t)]^n}{[1+v(t,t)]^{n+1}}\,[1-A(t)]^{n_0} \\ &\times \sum_{k=0}^{\min\{n_0,n\}} \binom{n_0}{k}\binom{n}{k}\left[\frac{1}{v(t,t)}\frac{A(t)}{1-A(t)}\right]^k, \end{aligned} \tag{20}$$

and $A(t) = \frac{|u(t,t_0)|^2}{1+v(t,t)}$. The density matrix $\rho(t)$ in Eq. (19) is diagonal in the energy eigenbasis $|n\rangle$, there is no coherence in energy eigenbasis. The function $p_n^{n_0}(t)$ is the probability of finding the quantum system in a particular energy eigenstate $|n\rangle$ at time $t$. In this case, the von Neumann entropy $S(\rho(t))$ and thermodynamic entropy $\mathcal{S}_{th}(t)$ will be the same, even during the nonequilibrium time evolution. There will be no contribution of coherence in the entropy balance equation. Hence, the thermodynamic entropy can simply be obtained in this case as

$$\mathcal{S}_{th}(t) = -\sum_{n=0}^{\infty} p_n^{n_0}(t) \ln p_n^{n_0}(t), \tag{21}$$

and the entropy $\mathcal{S}_{th}(t)$ can experimentally be measured through the probability distribution $p_n^{n_0}(t)$.

The average energy of the system at an arbitrary time, namely the nonequilibrium internal energy, is given by

$$U(t) = \hbar\,\omega(t)\;\mathrm{n}(t), \tag{22}$$

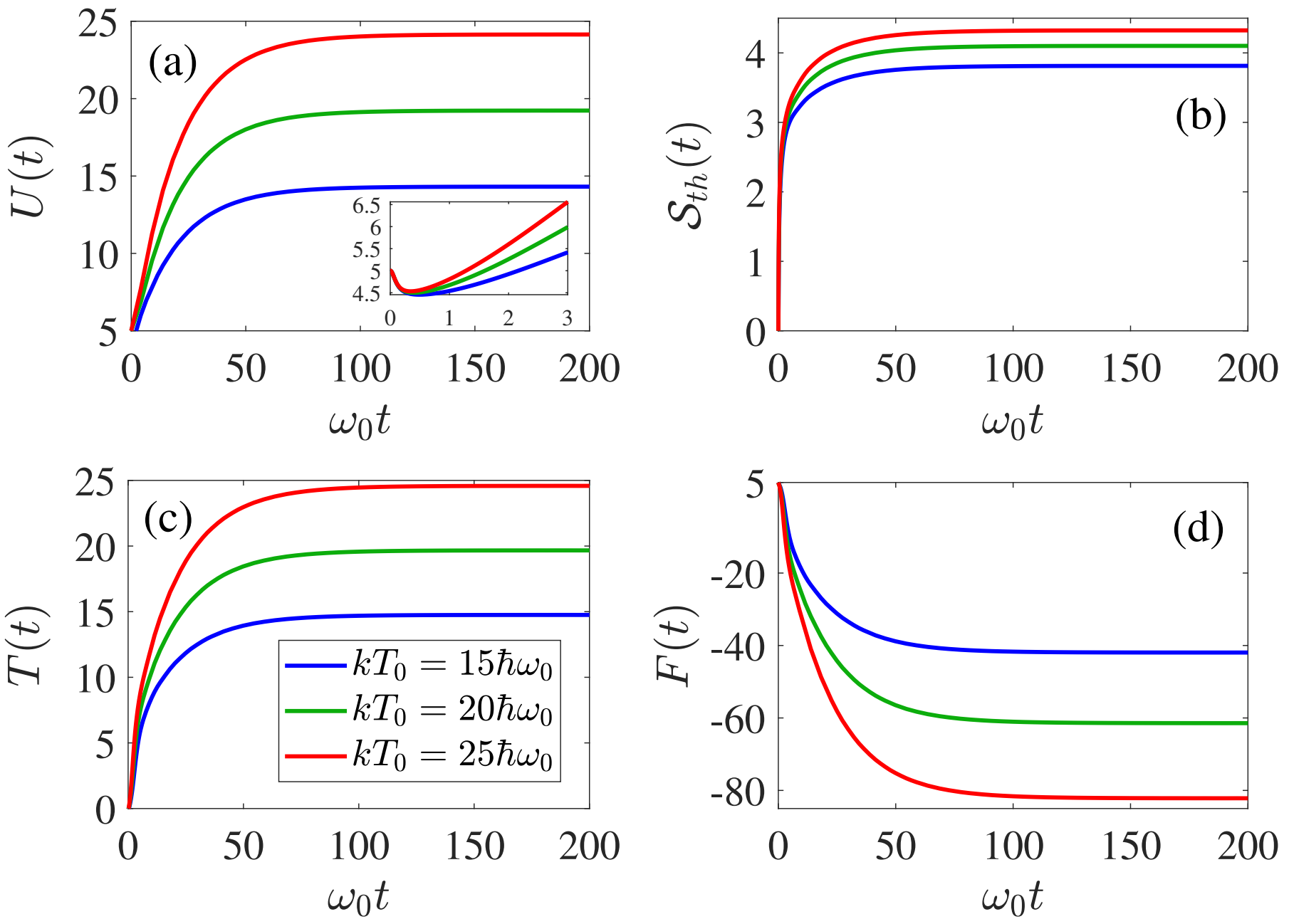

**Supplementary Figure S1:** Quantum thermodynamic quantities under nonequilibrium dynamics: Time evolution of (a) the internal energy $U(t)$, (b) the thermodynamic entropy $\mathcal{S}_{th}(t)$, (c) the dynamical temperature $T(t)$, and (d) the free energy $F(t)$. The system is initially prepared in the Fock state $|n_0\rangle = |5\rangle$. In each panel, the three curves correspond to different initial temperatures of the thermal reservoir: $kT_0 = 15\hbar\omega_0$ (blue), $20\hbar\omega_0$ (green), and $25\hbar\omega_0$ (red). The system-reservoir coupling strength is fixed at $\eta = 0.1\eta_c$, with a reservoir cutoff frequency $\omega_c = 10\omega_0$.

**Alt Text:** The time evolution of key nonequilibrium thermodynamic quantities for the open quantum system initially prepared in a Fock state for different initial reservoir temperatures. The four panels display the internal energy, thermodynamic entropy, dynamical temperature, and free energy, illustrating their evolution as the initially incoherent quantum system relaxes toward thermal equilibrium.

where $\mathrm{n}(t) = \mathrm{Tr}[a^\dagger a \rho(t)] = n_0\, |u(t,t_0)|^2 + v(t,t)$. Using the first law of nonequilibrium quantum thermodynamics, the rate of change of work for this system is given by

$$\dot{W}(t) = \hbar\, \dot{\omega}(t)\, \mathrm{n}(t), \tag{23}$$

and the rate of heat flow between the system and the reservoir is given by

$$\frac{dQ(t)}{dt} = \hbar\omega(t)\Big\{\widetilde{\gamma}(t) - 2\gamma(t)\mathrm{n}(t)\Big\} \tag{24}$$

From these results, one can determine the dynamical temperature $T(t)$ of the system given by

$$T(t) = \frac{dQ(t)}{dt} \bigg/ \frac{d\mathcal{S}_{th}(t)}{dt} \tag{25}$$

Consequently, the nonequilibrium free energy for this system is given by

$$F(t) = U(t) - T(t)\ \mathcal{S}_{th}(t), \tag{26}$$

We apply our nonequilibrium quantum thermodynamic framework to analyze an open quantum system initially prepared in the Fock state $|n_0\rangle = |5\rangle$. Supplementary Figs. S1a-S1d show the time evolution of the internal energy $U(t)$, thermodynamic entropy $\mathcal{S}_{th}(t)$, dynamical temperature $T(t)$, and nonequilibrium free energy $F(t)$ for weak system-reservoir coupling and for different initial temperatures ($T_0$) of the reservoir. As the system undergoes nonequilibrium quantum evolution, both $U(t)$ and $\mathcal{S}_{th}(t)$ increase monotonically (Supplementary Figs. S1a-S1b) and saturate at values set by $T_0$. The dynamical temperature $T(t)$ rises smoothly toward $T_0$ (Supplementary Fig. S1c), while the free energy decreases monotonically (Supplementary Fig. S1d). As the system undergoes nonequilibrium evolution, the internal energy and thermodynamic entropy increase monotonically before saturating at values determined by the initial reservoir temperature. The dynamical temperature of the system $T(t)$ evolves continuously toward the reservoir temperature $T_0$, while the nonequilibrium free energy $F(t)$ decreases monotonically to its minimum equilibrium value. These results demonstrate that the proposed definitions of thermodynamic entropy, dynamical temperature, and free energy consistently recover the expected equilibrium thermodynamic behavior in the absence of quantum coherence.

We also analyze the first law of thermodynamics, which relates the time derivative of the system's internal energy to the work rate and heat currents. Supplementary Figs. S2a-S2c display the temporal behavior of $\dot{U}(t)$, $\dot{W}(t)$, and $\dot{Q}(t)$ for various initial reservoir temperatures $T_0$ in a far-from-equilibrium regime. At early times, $\dot{U}(t)$ becomes negative, reflecting a transient decrease of the internal energy, consistent with the short-time behavior of $U(t)$ shown in the inset of Supplementary Fig. S1a. The negative work rate in Supplementary Fig. S2b indicates that work is performed on the system by the reservoir. The long-time thermalization dynamics are governed almost entirely by heat exchange between the system and the reservoir, while the work contribution is restricted to a short transient regime. As time progresses, both the work and heat rates decay to zero, signaling relaxation toward thermal equilibrium. We have checked this through numerical computation. Together, these results verify that the proposed definitions of work and heat remain fully consistent with the first law of thermodynamics even when the system evolves from an initially incoherent state, thereby demonstrating that the present framework is applicable to both coherent and incoherent nonequilibrium quantum dynamics.

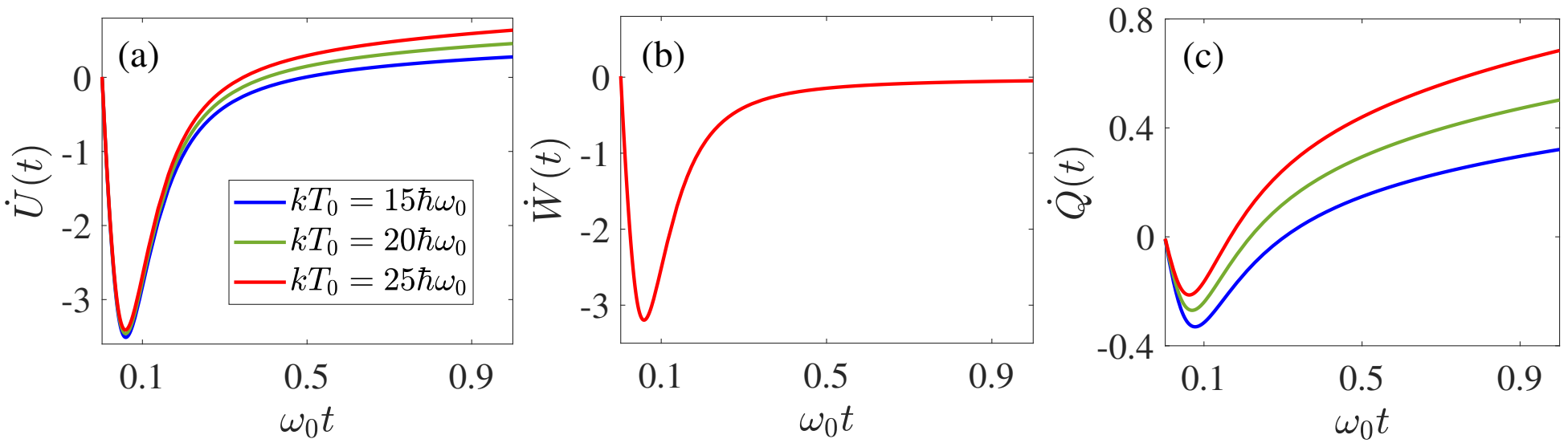

**Supplementary Figure S2:** Nonequilibrium dynamics of internal energy, work, and heat exchange: Time evolution of (a) the rate of change of internal energy $\dot{U}(t)$, (b) the work rate $\dot{W}(t)$, and (c) the heat flow rate $\dot{Q}(t)$, respectively. The system is initially prepared in the Fock state $|n_0\rangle = |5\rangle$. We consider three different initial temperatures of the thermal reservoir: $kT_0 = 15\hbar\omega_0$ (blue), $20\hbar\omega_0$ (green), and $25\hbar\omega_0$ (red). The work-rate dynamics exhibits negligible sensitivity to reservoir temperature $T_0$. The system-reservoir coupling strength is fixed at $\eta = 0.1\eta_c$, with a reservoir cutoff frequency $\omega_c = 10\omega_0$.

**Alt Text:** The time evolution of the rates of change of internal energy, work, and heat flow during the transient nonequilibrium dynamics of the open quantum system initially prepared in a Fock state for different initial reservoir temperatures. The three panels illustrate the short-time evolution of these quantities while the initially incoherent quantum system remains far from thermal equilibrium.